# Study of the Crosstalk Evaluation for Cavity BPM

Jian Chen(陈健)[1,2]    Yong-bin Leng(冷用斌)[1],*    Lu-yang Yu(俞路阳)[1]

Long-wei Lai(赖龙伟)[1]    Ren-xian Yuan(袁任贤)[1]

[1]Shanghai Institute of Applied Physics, Chinese Academy of Science, Shanghai 201204, China

[2]University of the Chinese Academy of Science, Beijing 100049, China

**Abstract:** In order to pursue high-precision beam position measurements for the free-electron laser (FEL) facilities, cavity beam position monitor (CBPM) is employed to measure the transverse position which can meet the requirement of position resolution with a sub-micrometer or even nanometer scale. But for the pill-box cavity BPM, the possible existed crosstalk between the cavities will have effects on the accurate measurement of beam position. Two methods, the principle component analysis (PCA) method and the method of harmonic analysis, are proposed in this paper to evaluate the crosstalk based on the experiment dates from the low quality CBPM prototype in Shanghai Deep ultraviolet free electron laser (SDUV-FEL) facility and high quality CBPM in Dalian Coherent Light Source (DCLS), respectively. The results demonstrated that these two methods are feasible in evaluating the crosstalk between the cavities.

## 1    Introduction

Free-electron laser based on the linear accelerator is the fourth generation light source which has the characters such as high brightness, short wavelength, full coherence, ultra-fast time resolution. So it has become an extremely important research equipment to meet the demands of biological, chemical, and material science research[1]. But for FEL facilities, to keep the interaction between the electron beams and generated photon beams, it is critical to ensure a tight overlap between them along a long undulator section. Comparing with the various types of BPMs such as button and stripline BPM, cavity BPM adopts resonant cavity structure and use the characteristic modes excited by the electron beam to measure the beam position has the advantage of high-resolution is widely used in the FEL facilities[2].

For the cylindrical pill-box CBPM, due to the inevitable fabrication tolerance will produce in the actual processing, the cavity will have a random deformation which will make a change of the polarization direction of the electromagnetic field and will have an effect on the accurate measurement of the beam position.

For the construction of the Shanghai soft X-ray free-electron laser (SXFEL) facility and Dalian Coherent Light Source (DCLS) facility[3,4], the low quality (Q) and high Q cavity BPM prototype were designed and developed by our group[5-9]. The relevant beam experiments were also performed on the SDUV and DCLS facility. To study the effect of crosstalk between cavities for the beam position measurement, a movement platform is introduced to control the transverse placement of the CBPM and a broadband oscilloscope was used to sample the RF signal from the cavity.

In this paper, we described the causes of the transverse crosstalk and proposed two methods to evaluate the crosstalk between cavities based on the RF signal of the CBPM probe. The specific analytical method will be discussed in the following part.

## 2    Detection principle of the cylindrical CBPM

For a cylindrical pill-box CBPM which can adopt a resonant cavity structure and through the use of

Supported by National Natural Science Foundation of China (No.11575282    No.11305253)

E-mail: chenjian@sinap.ac.cn

*Corresponding author. E-mail: lengyongbin@sinap.ac.cn

antisymmetric characteristic modes, coupled from the cavity, to measure the beam position. When the beam source runs along the z-axis, the axial electric field component of the TM110 mode in cylindrical coordinates can be expressed by Eq. (1):

$$E_z(\rho, \phi, z) = E_0 J_1(\frac{\chi_{11}\rho}{r})\cos\phi. \qquad (1)$$

where $E_0$ is the amplitude of the electric field and $J_1$ is the first-order Bessel function of the first kind, $\chi_{11}$ is the first root of $J_1(\rho) = 0$, $r$ is the cavity radius and $\rho$ is the radial coordinate. Since $J_1(\rho)$ is proportional to $\rho$ when $\rho \sim 0$. Therefore, the excited voltage of the TM110 mode is proportional to the beam offset x and beam charge q. which can be written by Eq. (2):

$$V_z = A_0 q x. \qquad (2)$$

From the Eq. (2), the RF voltage is zero when the beam at the cavity center and proportional to the beam offset. Thus, even a tiny variation can also be easily detected by using a high gain amplifier if the beam position is close to the center. But for the button and stripline BPM, when the electron beam close to the center, the signals from opposite electrodes needs to be subtracted which will result in a significant loss of the effective number of bits. It limits their resolution only about 10μm. This is the reason why the cavity BPM has the capability for the high resolution measurement.

In order to eliminate the variation effect of the beam charge, an additional monopole TM010 mode cavity is also employed. The signal amplitude of the monopole mode is independent to the beam position but is proportional to the beam charge only[10].

## 3    The causes of the Transverse crosstalk

Transverse crosstalk of the cavity includes the crosstalk between the reference cavity and the position cavity as well as the crosstalk between the horizontal position cavity and the vertical position cavity. If the distance between the reference cavity and the position cavity is short and has a large diameter of the pipe between the cavities, then there has the possibility of crosstalk between them. But this type of crosstalk is easy to be avoided in the process of the cavity design. Relatively speaking, the problem of crosstalk between horizontal and vertical cavity is more common. The ideal cylindrical cavity BPM as shown in Fig. 1 (left), the resonant modes polarized in the direction of displacement excited by the beam which offset the center axis (x,y) is equal to the sum of the two resonant modes polarized in the horizontal and vertical directions by the displacement x and y, respectively. The waveguides mounted in horizontal and vertical direction couple the signals so as to measure the correspondent displacement [11].

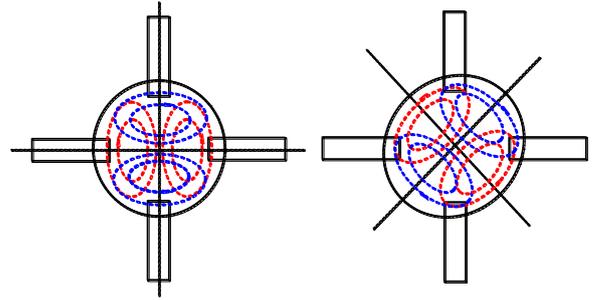

Fig.1. Ideal cavity and coupling (left) and cavity with distortion (right).

In reality, there are always has some small distortions of the cavity symmetry due to the fabrication tolerance, welding procedure and so on. These distortions can be treated that the cavity as slightly elliptical deformed but that cavity distortions caused by the fabrication which makes the orientation of the axis of the ellipse are basically unpredictable. As shown in Fig. 1 (right), because of the slightly distortion in cavity, the polarizations of the two excited dipole modes of the magnetic fields are perpendicular to each other but the polarization direction of TM110 modes no longer consistent with the mounting direction of the waveguide[12]. Under this circumstance, a set of waveguides couples two polarized resonant modes simultaneously to cause crosstalk.

In order to suppress the crosstalk, except to adopt the rectangular cavity structure, an effective approach is to introduce a structure which can make polarization directions are fixed on the X and Y axes. Explicit methods can be found in the reference [11-12]. Although crosstalk can be suppressed by the method describe above, it is still unavoidable for the cylindrical cavity. So it is essentially to evaluate the impact of transverse crosstalk on the measurement of the beam position.

## 4  PCA method

To meet the high resolution beam position measurement requirement for SXFEL and DCLS, a low Q CBPM was developed in the exploration process. An evaluation of the cavity performance with beam has been performed on the SDUV facility[7].

### 4.1  System setup

To research the position dependence and the crosstalk of the output signals from the cavities, a two-dimensional motion platform was installed under the cavity which can imitate the beam offset both in the horizontal and vertical directions. A broadband oscilloscope with 6 GHz bandwidth, 25 GHz sampling rate is also used as the data acquisition equipment. The diagram of the evaluation system is detailed in Fig. 2. With the motion platform, the other direction is set at 0 μm when moving in the one direction. The output signals from the reference cavity and position cavity in the every step are recorded in three channels of the oscilloscope simultaneously.

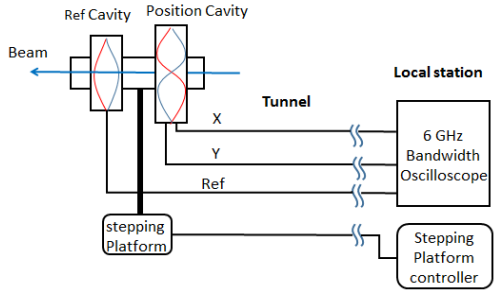

Fig.2. Diagram of the system.

### 4.2  Principle of the PCA method

Principal Component Analysis (PCA) is a commonly used method for data analysis that transforms raw data into a set of linearly independent representation of each dimension by linear transformation so that the main characteristic components of the data can be separated and extracted.

The electromagnetic field in a cavity or a signal sampled from the cavity is a linear combination of the orthogonal modes which determined by the property of the cavity. It can be expressed by the Eq. (3):

$$V(t) = \Sigma_{m,n,p} C_{m,n,p} V_{m,n,p}(t) + C_{noise} V_{noise}(t) . \quad (3)$$

where the $V_{m,n,p}(t)$ represents the temporary vectors for various orthogonal modes, so the time evolution properties such as the resonant frequency and the damping ratio of the modes can be studied separately and it also used to determine the physical source of the modes. The normalized $C_{m,n,p}$ represents the spatial vectors which indicates the variance in the amplitude during the measurements. The last term is considered to be a component at the noise level.

Based on the property of the cavity and a data set of broadband RF signal with different beam positions, the PCA method can be used to evaluate the performance including mode identification, crosstalk evaluation and eliminate unwanted coupling to improve the measurement accuracy and so on[13-16].

### 4.3  Mode identification and crosstalk evaluation

For the mode identification and crosstalk evaluation in vertical position cavity, move the cavity vertically and acquire the signals of vertical position cavity in every step to form a signal matrix. Using PCA on the signal matrix, the singular values of the modes were obtained which can be seen in Fig .3. There have three modes were higher than the noise floor obviously.

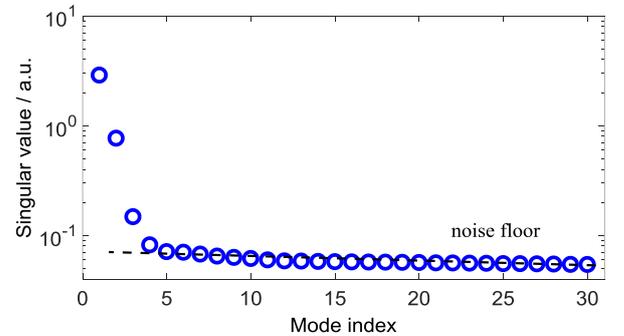

Fig.3. Singular values of the modes.

The strength of the first two modes is much higher than other modes. Based on the spectrum of the temporal vectors as shown in Fig. 4 (for ease of observation, the amplitude of modes have not multiplied by the correspondent singular value) that the resonant frequency are 4.77 GHz and the Q factor are 68 which consistent with the design value, and the mode amplitude of the first two modes are linear related to the vertical position of the beam which can be seen at the spatial vectors distribution of the first three modes (Fig .5). These indicate that the mode 1 and 2 are jointly characterizing the TM110 mode of the position cavity, that is, the main operation mode of the position cavity.

For the third mode that the mode amplitude is irrespective to the beam position, combine with the spectrum of the temporal vectors, this mode not only include the TM010 mode ($f_{pos,010}$ = 3.3 GHz) and the TM020 mode ($f_{pos,020}$ = 6.2 GHz) of the position cavity but also have a strong mode come from the reference cavity ($f_{ref,010}$ = 4.7 GHz, Q = 223).

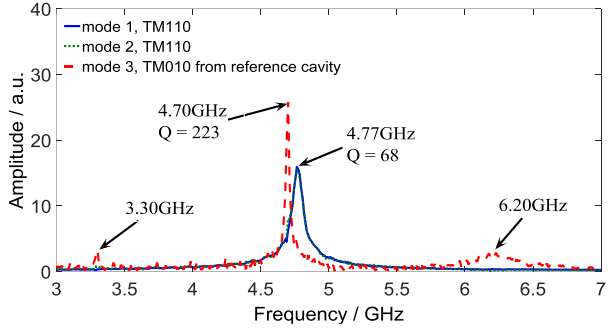

Fig.4. Temporal vectors distribution of the first three modes.

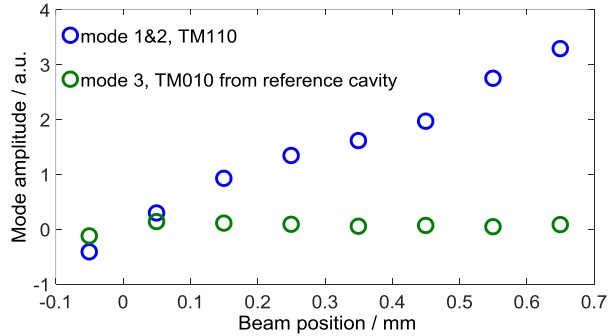

Fig.5. Spatial vectors distribution of the first three modes.

The existence of the crosstalk between reference cavity and vertical position cavity is detected by using the PCA method which is caused by the short distance between the two cavities (35 mm) so that the damping effect is not enough. Because of the resonant frequency of the reference cavity is closed to the position cavity, the effect of the crosstalk should be evaluated carefully.

The intensity of the principal mode signal (4.77 GHz) at different positions was obtained by harmonic analysis, as shown in Fig. 6, the position sensitivity of the principal mode was also got by the linear fitting. The crosstalk signal from the reference cavity is unrelated to the beam position so the amplitude is a fixed value which can be obtained by the separated mode. As illustrated in Fig. 7, combine with the position sensitivity, the effect of the crosstalk from the TM010 mode of the reference cavity to the TM110 mode of the vertical position cavity is about 4 μm.

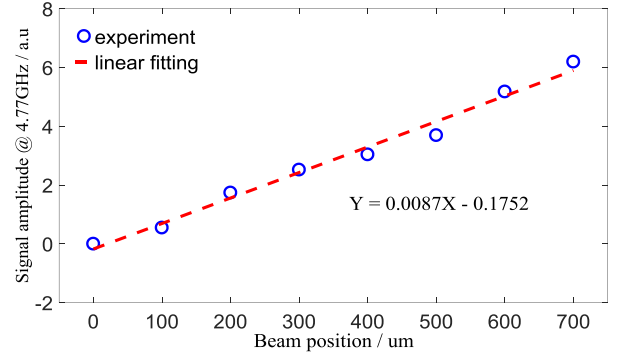

Fig.6. Linear relation between the beam position and the amplitude of the TM110 mode.

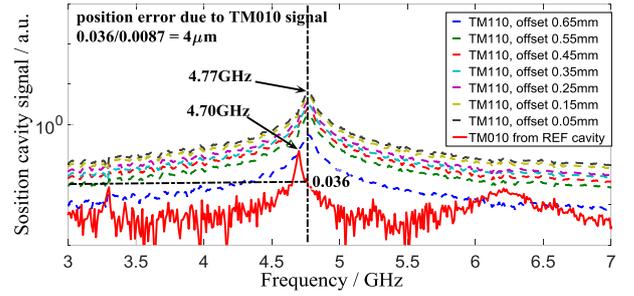

Fig.7. Amplitudes of the TM110 mode of the position cavity and the crosstalk mode from the TM010 mode of the reference cavity.

Based on the results analyzed above, we verified that the PCA method is a powerful tool to analyze the RF signal of the cavity probe. By moving the position of the probe to construct the data matrix and PCA is benefit to separate the two types of the resonant modes in the original RF signal, the position-independent mode and position-dependent mode, and combine with the information of the signal strength of the mode, resonant frequency and Q value that we can confirm the source of the main mode, measure the characteristic parameters and analyze the influence of the mode on position measurement.

## 5    Method of harmonic analysis

On the basis of the design experience of low Q cavity BPM, we have redesigned a high Q cavity BPM also for more in-depth study. The relevant parameters of the high Q cavity can be seen in Ref [8].

## 5.1 System setup

In order to evaluate the transverse crosstalk of the high Q CBPM with the beam, a tested system similar to the low Q CBPM, except a pre-amplifier front-end was added in the tunnel which can minimize the RF signal losses from the long connection cables, was built. The two-dimensional motion platform was also installed under the cavity to imitate the beam offset from -300 to 300 μm with the step of 100 μm. The diagram of the evaluation system is detailed in Fig. 8.

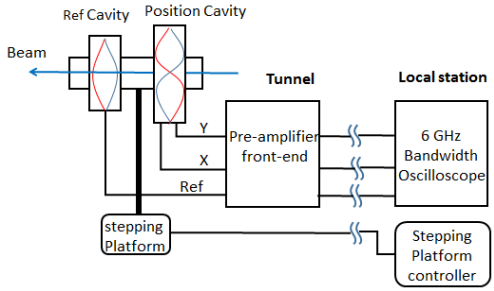

Fig.8. Diagram of the system.

## 5.2 Principle of the harmonic analysis

The characteristic of the signal spectrum of the cavity is dependent on the characteristics of the cavity only, such as the resonant frequency and Q value, and independent on the position of the beam in the cavity. These also can be illustrated by the Fourier transform[17] of the sinusoidal attenuation signals which shown in Eq. (4):

$$F(\omega) = \frac{1}{2*pi} * \frac{e^{i(\varphi - \omega_0 t_0)}}{\frac{1}{2\tau} + i(\omega - \omega_0)} . \qquad (4)$$

Where the $\omega_0$ is the resonant angular frequency, $\tau$ is the decay time of the cavity and $t_0$ is the initial time of the excitation of the cavity signal. The expression of the signal spectrum is take modulo for the Eq. (4), which can be written by Eq. (5):

$$F = \frac{A}{\sqrt{(\frac{1}{2\tau})^2 + 4*pi^2*(f - f_0)^2}} . \qquad (5)$$

Where the *A* is a variable which only dependent on the beam offset. The $\tau$ and $f_0$ are constants for a given cavity which can be fitted by the Eq. (5) based on the data of the cavity. It is better to fit at a bigger beam offset so as to diminish the impact of the crosstalk.

When the spectrums of the horizontal and vertical direction are fitted respectively in the bigger offset to obtain the correspondent $\tau$ and $f_0$, then the overall fitting function is obtained as shown in Eq. (6):

$$F_{a\,l\,l} = A*_x F +_x A*_y F. \qquad (6)$$

$A_x$ and $A_y$ represent the signal magnitude of different cavities, so that the signals of different cavities can be separated for analysis.

## 5.3 Analysis of the crosstalk

Based on the experimental data and the method of harmonic analysis, because the reference cavity almost none crosstalk to the position cavity which can be seen in Fig. 9, we only evaluated the crosstalk of the CBPM between horizontal and vertical cavity.

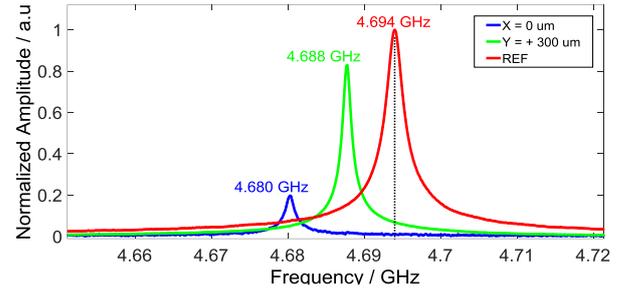

Fig.9. The frequency spectrum of the reference and position cavity.

Moving the cavity horizontally and fitting the data of vertical channel with the Eq. (6) at every position. Fig. 11 illustrated the spectrum of overall fitting when the horizontal is located at +300 μm and vertical is located at 0 μm, combine with the calibration factor of the vertical direction (Fig. 10), we can see that the crosstalk of the horizontal position cavity to the vertical position cavity is about 0.84 μm.

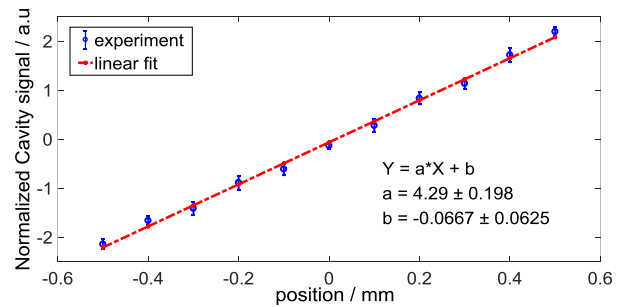

Fig.10. Calibration factor of the vertical position cavity.

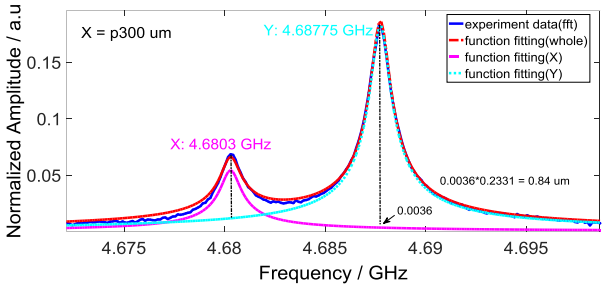

Fig.11. Crosstalk from horizontal cavity to vertical cavity when x = p300μm.

The degree of the crosstalk from the horizontal position cavity to the vertical position cavity can be obtained by the slope as shown in Fig. 12, about -52.6 dB.

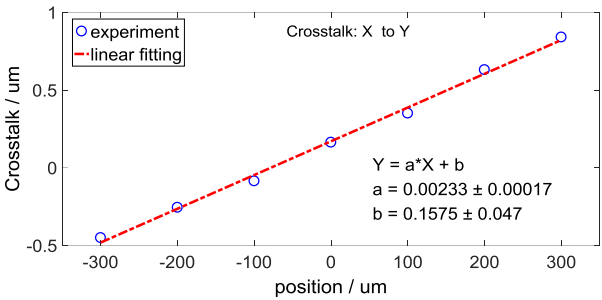

Fig.12. Degree of crosstalk from the horizontal position cavity to the vertical position cavity.

The crosstalk from the vertical position cavity to horizontal position cavity is also handled the same way. The Fig. 13 and 14 illustrated that the crosstalk of the vertical position cavity to the horizontal position cavity is about 0.16μm when the vertical is located at +300 μm. The degree of the crosstalk is also obtained as shown in Fig. 15, which is smaller compared with the crosstalk from the horizontal position cavity to the vertical position cavity, about -71.4 dB.

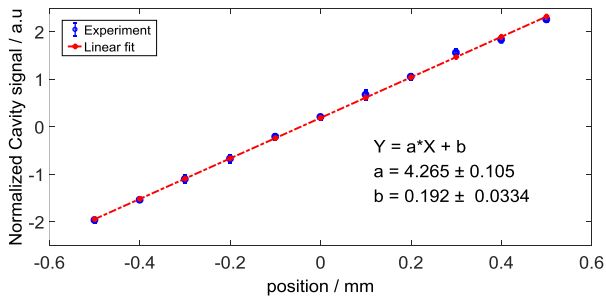

Fig.13. Calibration factor of the horizontal position

cavity.

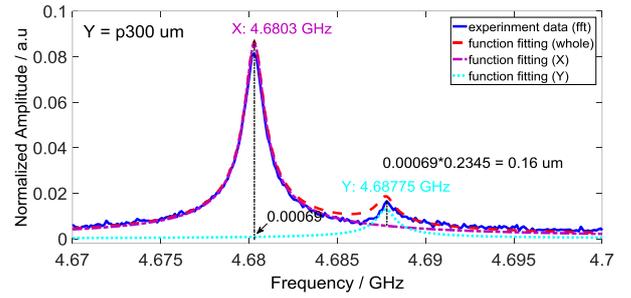

Fig.14. Crosstalk from vertical cavity to horizontal cavity when y = p300μm.

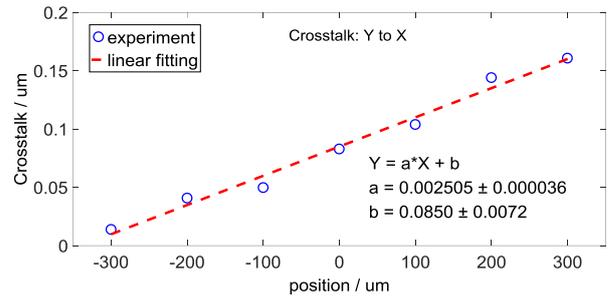

Fig.15. Degree of crosstalk from the vertical position cavity to the horizontal position cavity.

From the analysis above, we can see that the method of harmonic analysis is very effective in the analysis of the crosstalk between the cavities. The results shown that the isolation between the cavities to meet the requirement of less than -40 dB. However, this method also has limitation in analyzing crosstalk problem. When the Q value and the resonant frequency of the analyzed cavities are the same, the method of harmonic analysis cannot fit and separate the signals effectively.

## 6   Conclusion

In this paper, we described the causes of the crosstalk and proposed two methods to evaluate the crosstalk between cavities. The PCA method decomposes the RF signal matrix directly from the cavity and the position-independent mode and position-dependent mode can be separated, combine with the characters of the cavity, the influence of the mode on position measurement can be analyzed. Applied to the low Q prototype cavity, the effect of the crosstalk from the reference cavity to the vertical position cavity about 4 μm was evaluated. The method of harmonic analysis by the way of fitting the spectrum of the cavity signal in the frequency domain based on the cavity characteristics, which can analyze more conveniently and clearly but it also limited by the resonant frequency of the cavities to be analyzed are different and better used in the cause of higher Q value. Combined with the experiment we have done in DCLS, the crosstalk between horizontal and vertical position cavities are less than -50 dB. These shown that the two methods are effectively and practicably to analyze the crosstalk between cavities.

# 7    Acknowledgements

The authors are grateful to Meng Zhang, Dazhang Huang and members of DCLS for providing a better test platform as well as the work of adjusting the beam status. This work was partially supported the National Natural Science Foundation of China (11575282 and 11305253).